\documentclass[11pt,a4paper]{article}

\usepackage{amsmath,amssymb,amsfonts,bm}
\usepackage{graphicx}
\usepackage[margin=2.5cm]{geometry}
\usepackage{booktabs}
\usepackage{multirow}
\usepackage{siunitx}
\usepackage{subcaption}
\usepackage{mathtools}
\usepackage{cite}
\usepackage{xcolor}
\usepackage{url}
\usepackage[hidelinks]{hyperref}

\graphicspath{{figures/}}

\title{Joint Retrieval of Radial Wind, Terminal Fall Velocity, and Median Diameter From Single-Polarization Fast-Scanning Weather Radar}

\author{Tworit~K.~Dash, Hans~Driessen, Oleg~A.~Krasnov, and Alexander~G.~Yarovoy}
\date{}

\newenvironment{IEEEkeywords}
  {\par\smallskip\noindent\textbf{Keywords: }}
  {\par\smallskip}

\begin{document}

\maketitle

\begin{abstract}
The inverse problem of precipitation retrieval from Doppler power spectral density (PSD) measurements of a fast-scanning single-polarization X-band weather radar is addressed. The formulation is developed around the quantities that remain most interpretable and stable across short coherent processing intervals (CPIs): radial-wind mean, radial-wind spectral width, reflectivity-weighted terminal fall velocity, and median volume diameter. The gamma drop-size distribution (DSD) shape and rate parameters $(\eta,\Lambda)$ are retained as latent variables within the spectral forward model, with the intercept parameter $N_0$ treated separately. This choice is motivated by likelihood analyses showing that the inverse problem is weakly identifiable with respect to the latent DSD pair, whereas the derived quantities remain substantially more stable. Simulated-data experiments compare the proposed approach with a moment-based retrieval using externally supplied wind information in the short-CPI regime. Real-data experiments demonstrate same-scan spatial likelihood pooling through a one-scan plan-position-indicator (PPI) retrieval and 33-scan time-series comparisons against two disdrometers within the radar coverage region. The results show that the latent DSD parameters should be treated as intermediate fitting variables rather than primary end products. The proposed approach shows its clearest practical benefit for the median volume diameter $(D_m)$ and reflectivity-weighted terminal fall velocity $(V_T)$ under incoherent short-CPI operation. Spatial pooling allows efficient mapping of $D_m$ and $V_T$ over a full PPI when each Doppler spectrum contains only a few slow-time samples. Radar--disdrometer discrepancies remain physically meaningful even when the spectral fit and temporal evolution are reasonable.
\end{abstract}


\section{Introduction}
Fast-scanning weather radars operate under a measurement constraint that is central to this paper: short dwell times imply short coherent processing intervals (CPIs), and short CPIs produce noisy Doppler power spectral density (PSD) estimates when each look is processed in isolation. This is particularly relevant for compact X-band scanning radars used for rapid-update precipitation monitoring, gap filling, and urban hydrometeorological sensing. In those settings the data available to the retrieval algorithm are often several incoherent short records, gathered either from neighboring radar cells or from successive scans, rather than one long coherent time series \cite{Dash2024,Dash2022PerformanceRadar}. The motivation is especially strong for aviation-weather applications, where timely updates on precipitation structure and wind conditions are operationally important \cite{OudeNijhuis2018WindProject}.

The Doppler PSD in rain is shaped by both air motion and the drop-size distribution (DSD). The wind field shifts and broadens the spectrum, while the raindrop sizes determine the distribution of terminal fall speeds and therefore the vertical-fall contribution to the spectrum. In the model used here, the precipitation part is described with a gamma DSD. That DSD does not appear in the retrieval as an abstract microphysical add-on; it enters directly through the fall-speed part of the Doppler spectrum. In plain terms, different drop-size populations produce different fall-speed populations, and those fall-speed populations leave a measurable imprint on the PSD.

This is also where the main difficulty enters. A short-CPI single-polarization spectrum does contain information about the precipitation population, but that information is not distributed evenly across all DSD parameters. The same measured PSD can often be explained by several nearby combinations of wind broadening and gamma-DSD shape. As a result, some parameter combinations are much better constrained than others.

The single-polarization Doppler spectrum is therefore not equally informative about all parameters of a gamma DSD. In practice, the latent pair $(\eta,\Lambda)$ can move substantially while producing very similar fitted spectra, especially when wind broadening is uncertain. This observation does not invalidate the model; rather, it determines which quantities can be reported most credibly as primary retrieval products.

Accordingly, the paper focuses on reflectivity-weighted terminal fall velocity and median diameter, together with the radial-wind parameters required to separate the precipitation and wind contributions. The gamma-DSD parameters remain inside the forward model because they are needed to synthesize the spectrum, but they are not promoted as final geophysical outputs.

In a fast-scanning X-band weather radar, the return is not expected to contain a distinct zero-Doppler clear-air component of the type often exploited by VHF/UHF Bragg-scattering profilers \cite{Wakasugi1986ASpectra,Rajopadhyaya1993MeasuringRadar,Gossard1988MeasuringRadar,Williams2009RaindropStatistics,Currier1992CombinedDistributions}. X-band sensitivity to Bragg scattering is much weaker, and the measured precipitation spectrum therefore does not usually admit a clean wind-removal step based on a near-zero clear-air peak. The wind contribution and the precipitation contribution must instead be handled jointly in the forward model.

The proposed method is complementary to dual-polarization retrievals. Dual-polarization observables remain richer in microphysical information and are preferred when available and well calibrated, even though calibration and sensitivity issues remain important in practice \cite{Gatidis2022SensitivityRelationship}. In the remainder of the paper, the proposed method is referred to as single-polarization likelihood inversion technique (SPLIT). SPLIT addresses a different use case: single-polarization Doppler spectra from fast-scanning X-band radars, where several incoherent short-CPI measurements can be pooled across space or time.

The paper makes the following contributions.
\begin{enumerate}
    \item It reformulates the single-polarization inverse problem around radial-wind parameters, terminal fall velocity, and median diameter, with the gamma-DSD parameters treated as latent variables.
    \item It shows, through log-likelihood analysis, why the latent DSD parameters are weakly identifiable even when the fitted spectrum is accurate.
    \item It compares SPLIT with Chen-type \cite{Chen2020VerticalDistribution} equations in simulations using incoherent short-CPI records, with $N_0$ assumed known throughout all experiments.
    \item It demonstrates same-scan spatial likelihood pooling in real radar data, including one-scan plan-position-indicator (PPI) retrievals and 33-scan comparisons with two different disdrometers within the radar's maximum range.
\end{enumerate}

The main body of the paper is organized as follows. Section~II formulates the latent-variable forward model and the reported radar products. Section~III summarizes the likelihood-based retrieval. Section~IV examines the identifiability of the latent DSD pair. Section~V presents the simulation study, including the wind-broadening, incoherent-aggregation, and elevation-angle analyses. Section~VI presents the real-data experiments with one-scan PPI retrievals and site-based time-series comparisons. Section~VII discusses the implications of the results, and Section~VIII concludes the paper.

\section{Problem Formulation}
This section defines the forward model used throughout the paper and clarifies which quantities are treated as internal fitting variables and which are treated as reported radar products. The basic idea is simple: the measured Doppler spectrum is modeled as the combined effect of radial wind and precipitation fall speeds. The gamma DSD enters through the precipitation part of that spectrum, but the main outputs of interest are the wind quantities together with the median volume diameter $(D_m)$ and the reflectivity-weighted terminal fall velocity $(V_T)$, rather than the latent DSD pair itself.

\subsection{Latent and Reported Variables}
The forward model uses the gamma DSD
\begin{equation}
N(D)=N_0 D^\eta \exp(-\Lambda D),
\end{equation}
with known intercept parameter $N_0$ and latent shape--slope pair $(\eta,\Lambda)$. The gamma model is the most common analytical DSD model used in radar precipitation studies \cite{Ulbrich1983NaturalDistribution,Chen2020VerticalDistribution}. The unknown parameter vector is written as
\begin{equation}
\bm{\theta}=[\mu_v,\sigma_v,\eta,\Lambda]^\top ,
\end{equation}
where $\mu_v$ and $\sigma_v$ describe the radial-wind contribution and $(\eta,\Lambda)$ determine the precipitation spectrum shape.

The paper reports the derived quantities
\begin{equation}
D_m=\frac{3.67+\eta}{\Lambda},
\label{eq:Dm}
\end{equation}
and
\begin{equation}
V_T=\frac{C_1-C_2\left(1+\frac{C_3}{\Lambda}\right)^{-\eta-7}}{\sin\theta},
\label{eq:VT}
\end{equation}
instead of maps or sweeps of $\eta$ and $\Lambda$. In \eqref{eq:VT}, the constants $C_1$, $C_2$, and $C_3$ are inherited from the exponential raindrop fall-speed parameterization widely used in radar precipitation studies \cite{Atlas1973DopplerIncidence,Gunn1949TheAir}.

The latent pair $(\eta,\Lambda)$ is retained because it provides a compact and physically interpretable parameterization of the precipitation contribution to the Doppler spectrum. However, the principal geophysical products considered here are $\mu_v$, $\sigma_v$, $D_m$, and $V_T$, because these are both more interpretable and more stable under the single-polarization short-CPI measurement regime considered here.

\subsection{Single-Polarization PSD Model}
Let $Z_\ell(v_i)$ be the Doppler PSD at velocity bin $v_i$ for the $\ell$th incoherent record, with $\ell=1,\dots,L$. The expected spectrum is denoted by $F(v_i;\bm{\theta})$. For finite CPI length $N$, the semi-analytical expected PSD can be written as a generalized form of \cite[Eq.~(10)]{Dash2024},
\begin{align}
F(v;\bm{\theta}) &=
R\Bigg[
1+\sum_{q=1}^{N-1}\left(1-\frac{q}{N}\right)
\Big(
 Y(q)G(q)e^{-jq\zeta v}
\nonumber\\
&\qquad\qquad\qquad\qquad\qquad
 +Y(-q)G(-q)e^{jq\zeta v}
\Big)
\Bigg],
\label{eq:ModelSpectra}
\end{align}
where $R$ is the reflectivity term, $\zeta=4\pi T/\lambda$, $Y(q)$ is the normalized covariance associated with the radial-wind contribution, and $G(q)$ is the normalized covariance associated with the reflectivity-weighted vertical-fall contribution.

The key modeling step is that the radial velocity of a raindrop is treated as the sum of the radial wind component and the elevation-projected terminal-fall component. In the Doppler domain this means that the wind-broadened PSD and the fall-velocity PSD are convolved. Equivalently, in the lag domain their normalized covariance functions multiply, following the same argument used in \cite[Appendix~A]{Dash2024}. With the Gaussian wind model,
\begin{equation}
Y(q)=\exp\left(-\frac{1}{2}(\zeta q)^2\sigma_v^2\right)\exp\left(j\zeta q\mu_v\right).
\label{eq:Yq}
\end{equation}
For the precipitation contribution,
\begin{equation}
G(q)=\int_{0}^{\infty}\frac{N(D)D^6}{R}\frac{dD}{dV_{T\psi}}
\exp\!\left(jqV_{T\psi}\right)dV_{T\psi},
\label{eq:Gq1}
\end{equation}
with
\begin{equation}
R=\int_{0}^{\infty}N(D)D^6\,dD.
\label{eq:reflectivity_generic}
\end{equation}
Using the exponential fall-speed model and the same parameterization as in the earlier work, \eqref{eq:Gq1} becomes
\begin{align}
G(q) &=
\frac{e^{jqC_1}}{\Gamma(\eta+7)}
\int_{0}^{\infty}
y^{\eta+6}e^{-y}f(y)\,dy,
\label{eq:Gq2}
\end{align}
after the change of variables $y=\Lambda D$, where
\begin{equation}
f(y)=\exp\left(-jqC_2\exp\left(-\frac{C_3}{\Lambda}y\right)\right).
\label{eq:fy}
\end{equation}
The integral in \eqref{eq:Gq2} is evaluated using generalized Gauss--Laguerre quadrature,
\begin{equation}
G(q)\approx e^{jqC_1}\sum_{i=1}^{n}w_i f(y_i),
\label{eq:GLQ}
\end{equation}
where $y_i$ and $w_i$ are the quadrature nodes and weights. In this study $n=64$ is retained, and the range of $n$ for which the approximation remains reliable is discussed through the decay study in Appendix~A. This semi-analytical model is not only a generic PSD fit, but an explicit convolution-based model of wind broadening and vertical fall.

The use of several incoherent records is essential. Each individual Doppler spectrum is short and noisy, but the aggregate likelihood over multiple spectra retains the detailed PSD shape information that is lost in moment-only methods. This is the main regime in which SPLIT is intended to operate \cite{Dash2024,Dash2023PrecipitationPrior}.

\subsection{Why SPLIT Is an Approximate Method}
SPLIT remains an approximate technique. Its validity depends on the finite-$N$ approximation used to connect the time-domain lag structure to the Doppler PSD under the assumed gamma-law kernel. The approximation-validity study is therefore retained in Appendix~A, where the relevant decay scale is compared with the CPI length. The main text only states the operational implication: SPLIT should be trusted in the regime where the effective lag-domain support remains short relative to the available sequence length.

That approximation issue is particularly relevant for this paper because the method is deliberately tested in a short-CPI operating regime. The approximation is therefore not a side detail; it defines the domain in which the method is expected to be valid. For clarity, the appendices retain the theoretical checks, while the main paper concentrates on the retrieval implications.

\section{Likelihood-Based Retrieval}
Assuming independent exponential periodogram samples conditioned on the mean PSD, the log-likelihood for $L$ incoherent spectra follows the model used in \cite[Eq.~(12)]{Dash2024}:
\begin{align}
\log p(\mathbf{Z}\mid\bm{\theta})
&=
-\sum_{i=1}^{N_f}
\left[
L\log \left(\pi\left(F(v_i;\bm{\theta})+\sigma_n^2\right)\right)
\right. \nonumber\\
&\left.\qquad\qquad
+\frac{\sum_{\ell=1}^{L}Z_\ell(v_i)}{F(v_i;\bm{\theta})+\sigma_n^2}
\right],
\label{eq:ll}
\end{align}
where $\sigma_n^2$ is the noise power and $N_f$ is the number of retained velocity bins after removing the near-zero bins used only to avoid numerical sensitivity.

The SPLIT estimate is obtained by
\begin{equation}
\hat{\bm{\theta}}=\arg\max_{\bm{\theta}}\log p(\mathbf{Z}\mid\bm{\theta}).
\end{equation}
The reported outputs are then $\hat{\mu}_v$, $\hat{\sigma}_v$, $\hat{D}_m$, and $\hat{V}_T$, with $\hat{D}_m$ and $\hat{V}_T$ computed through \eqref{eq:Dm} and \eqref{eq:VT}.

The optimizer still searches in the four-dimensional latent space, but the four coordinates are not equally observable or equally useful as end products.

All synthetic experiments in this paper assume that $N_0$ is known. This assumption is intentional. It isolates the main retrieval problem of interest, namely the joint handling of wind and latent spectrum shape in the single-polarization PSD, without introducing an additional amplitude-identification ambiguity.

\section{Identifiability of the Latent DSD Pair}
The key scientific argument behind the paper is that the likelihood is much more informative about some combinations of parameters than about others. In particular, modest changes in the assumed wind broadening can induce large shifts in the best-fitting $(\eta,\Lambda)$ pair while leaving the fitted spectrum, the retrieved $V_T$, and the retrieved $D_m$ comparatively stable.

This behavior can also be expressed analytically through the local curvature of the log-likelihood. A second-order Taylor expansion of the log-likelihood around a reference point $\bm{\theta}_0$ gives \cite{vanderVaart1998AsymptoticStatistics}
\begin{align}
\log p(\mathbf{Z}\mid\bm{\theta}_0+\delta\bm{\theta})
&\approx
\log p(\mathbf{Z}\mid\bm{\theta}_0)
+\delta\bm{\theta}^{\top}\nabla_{\bm{\theta}}\log p(\mathbf{Z}\mid\bm{\theta})\big|_{\bm{\theta}_0}
\nonumber\\
&\qquad
+\frac{1}{2}\delta\bm{\theta}^{\top}\mathbf{H}(\bm{\theta}_0)\delta\bm{\theta},
\label{eq:ll_taylor_full}
\end{align}
where $\mathbf{H}(\bm{\theta}_0)$ is the Hessian matrix of the log-likelihood. If $\bm{\theta}_0$ is taken as a local maximizer, then the score term vanishes,
\begin{equation}
\nabla_{\bm{\theta}}\log p(\mathbf{Z}\mid\bm{\theta})\big|_{\bm{\theta}_0}=\mathbf{0},
\end{equation}
and the expansion reduces to
\begin{equation}
\log p(\mathbf{Z}\mid\bm{\theta}_0+\delta\bm{\theta})
\approx
\log p(\mathbf{Z}\mid\bm{\theta}_0)
-\frac{1}{2}\delta\bm{\theta}^{\top}\mathbf{I}(\bm{\theta}_0)\delta\bm{\theta},
\label{eq:ll_quadratic}
\end{equation}
where $\mathbf{I}(\bm{\theta}_0)\approx-\mathbf{H}(\bm{\theta}_0)$ is the local observed-information curvature and coincides with the usual Fisher-information form under the regular approximation used here. If $\delta\bm{\theta}$ lies close to an eigenvector associated with a small eigenvalue of $\mathbf{I}$, then the likelihood changes only weakly even for visibly different parameter values. In the present problem, that weak-curvature direction is dominated by the latent DSD coordinates. Therefore one can move $(\eta,\Lambda)$ away from the truth while producing only a negligible change in the fitted likelihood.

This link to the CRB is important. Large CRB variances for $(\eta,\Lambda)$ are not merely an asymptotic artifact here; they are the analytical counterpart of the broad ridge seen in the plotted likelihood slices. The data do not penalize motion strongly enough along that direction. By contrast, the reported quantities $D_m$ and $V_T$ vary more slowly along the same ridge, which is why they remain more stable than the latent pair itself even when the spectrum fit changes very little.

The weak curvature can also be seen directly from the reflectivity-weighted gamma law that underlies the precipitation part of the model. Because the spectrum is weighted by $D^6$, the relevant diameter distribution inside $G(q)$ is proportional to
\begin{equation}
p_6(D\mid \eta,\Lambda)\propto D^{\eta+6}e^{-\Lambda D},
\end{equation}
which is a gamma density with shape $\eta+7$ and rate $\Lambda$. In what follows, the subscript in $\mathbb{E}_6[\cdot]$ denotes expectation under this reflectivity-weighted density, namely
\begin{equation}
\mathbb{E}_6[h(D)] = \int_0^\infty h(D)\,p_6(D\mid \eta,\Lambda)\,dD.
\end{equation}
Therefore, for any $a>0$,
\begin{equation}
\mathbb{E}_6\!\left[e^{-aD}\right]
=
\left(1+\frac{a}{\Lambda}\right)^{-(\eta+7)}.
\label{eq:moment_collapse}
\end{equation}
This follows because the reflectivity weighting by $D^6$ converts the original gamma DSD into the normalized density
\begin{equation}
p_6(D\mid\eta,\Lambda)=
\frac{D^6N(D)}{\int_0^\infty D^6N(D)\,dD}
=
\frac{\Lambda^{\eta+7}}{\Gamma(\eta+7)}D^{\eta+6}e^{-\Lambda D},
\end{equation}
which is itself a gamma density with shape $\eta+7$ and rate $\Lambda$. Therefore
\begin{equation}
\mathbb{E}_6\!\left[e^{-aD}\right]
=
\frac{\Lambda^{\eta+7}}{\Gamma(\eta+7)}
\int_0^\infty D^{\eta+6}e^{-(\Lambda+a)D}\,dD,
\end{equation}
and the standard gamma-integral identity
\begin{equation}
\int_0^\infty x^{\nu-1}e^{-bx}\,dx=\frac{\Gamma(\nu)}{b^\nu}, \qquad b>0,
\end{equation}
gives
\begin{equation}
\mathbb{E}_6\!\left[e^{-aD}\right]
=
\frac{\Lambda^{\eta+7}}{(\Lambda+a)^{\eta+7}}
=
\left(1+\frac{a}{\Lambda}\right)^{-(\eta+7)}.
\end{equation}
Since the projected terminal-fall model is written as $V_{T\psi}=C_1-C_2 e^{-C_3 D}$, the reflectivity-weighted exponential moments that enter the semianalytical PSD are built from terms of the form
\begin{equation}
m_k(\eta,\Lambda)
=
\mathbb{E}_6\!\left[e^{-kC_3D}\right]
=
\left(1+\frac{kC_3}{\Lambda}\right)^{-(\eta+7)}, \qquad k=1,2,\dots
\label{eq:mk}
\end{equation}
For the parameter range of interest, $C_3/\Lambda$ is moderate to small, and
\begin{equation}
m_k(\eta,\Lambda)
\approx
\exp\!\left[-kC_3\frac{\eta+7}{\Lambda}\right].
\label{eq:mk_exp}
\end{equation}
Thus, to first order, the precipitation spectrum depends on $(\eta,\Lambda)$ mainly through the single combination
\begin{equation}
\rho=\frac{\eta+7}{\Lambda}.
\label{eq:rho_combo}
\end{equation}
If the spectrum is written schematically as $F(v;\eta,\Lambda)\approx \widetilde{F}(v;\rho)$ after fixing the wind parameters, then
\begin{equation}
\frac{\partial F}{\partial \eta}
\approx
\frac{1}{\Lambda}\frac{\partial \widetilde{F}}{\partial \rho},
\qquad
\frac{\partial F}{\partial \Lambda}
\approx
-\frac{\rho}{\Lambda}\frac{\partial \widetilde{F}}{\partial \rho},
\label{eq:collinear_gradients}
\end{equation}
so the two gradient components are approximately collinear. The $(\eta,\Lambda)$ Fisher submatrix therefore approaches
\begin{equation}
\mathbf{I}_{\eta\Lambda}
\approx
\alpha
\begin{bmatrix}
1\\[1mm]
-\rho
\end{bmatrix}
\begin{bmatrix}
1 & -\rho
\end{bmatrix},
\label{eq:fisher_rank1}
\end{equation}
for some nonnegative scalar $\alpha$, which is rank one. The weak-curvature direction is the direction that leaves $\rho$ nearly unchanged, i.e.
\begin{equation}
\delta \rho
=
\frac{\Lambda\,\delta\eta-(\eta+7)\delta\Lambda}{\Lambda^2}
\approx 0
\quad \Longrightarrow \quad
\delta\Lambda \approx \frac{\Lambda}{\eta+7}\,\delta\eta.
\label{eq:ridge_direction}
\end{equation}
This is the analytical counterpart of the ridge seen in the numerical likelihood slices.

The same approximation also explains why the derived quantities remain more stable. From \eqref{eq:VT},
\begin{equation}
V_T
=
\frac{C_1-C_2\left(1+\frac{C_3}{\Lambda}\right)^{-(\eta+7)}}{\sin\theta}
\approx
\frac{C_1-C_2e^{-C_3\rho}}{\sin\theta},
\label{eq:VT_rho}
\end{equation}
so $V_T$ is nearly constant along the ridge. Meanwhile,
\begin{equation}
D_m=\frac{\eta+3.67}{\Lambda},
\end{equation}
and along $\delta\rho\approx 0$,
\begin{equation}
\frac{dD_m}{d\eta}\bigg|_{\rho}
=
\frac{1}{\Lambda}
-\frac{\eta+3.67}{\Lambda^2}\frac{d\Lambda}{d\eta}
\approx
\frac{3.33}{\Lambda(\eta+7)},
\label{eq:Dm_ridge_slope}
\end{equation}
which is much smaller than the unit slope that would arise if $\Lambda$ were fixed. Hence the latent pair can move substantially while the spectrum, $V_T$, and even $D_m$ change much more slowly. The full model is not exactly rank deficient because the reflectivity factor and the higher-order terms in \eqref{eq:mk_exp} restore some curvature, but the approximation captures why the problem is intrinsically weakly conditioned in $(\eta,\Lambda)$.

\begin{figure*}[!t]
\centering
\includegraphics[width=\linewidth,trim=0.08cm 0.08cm 0.08cm 0.08cm,clip]{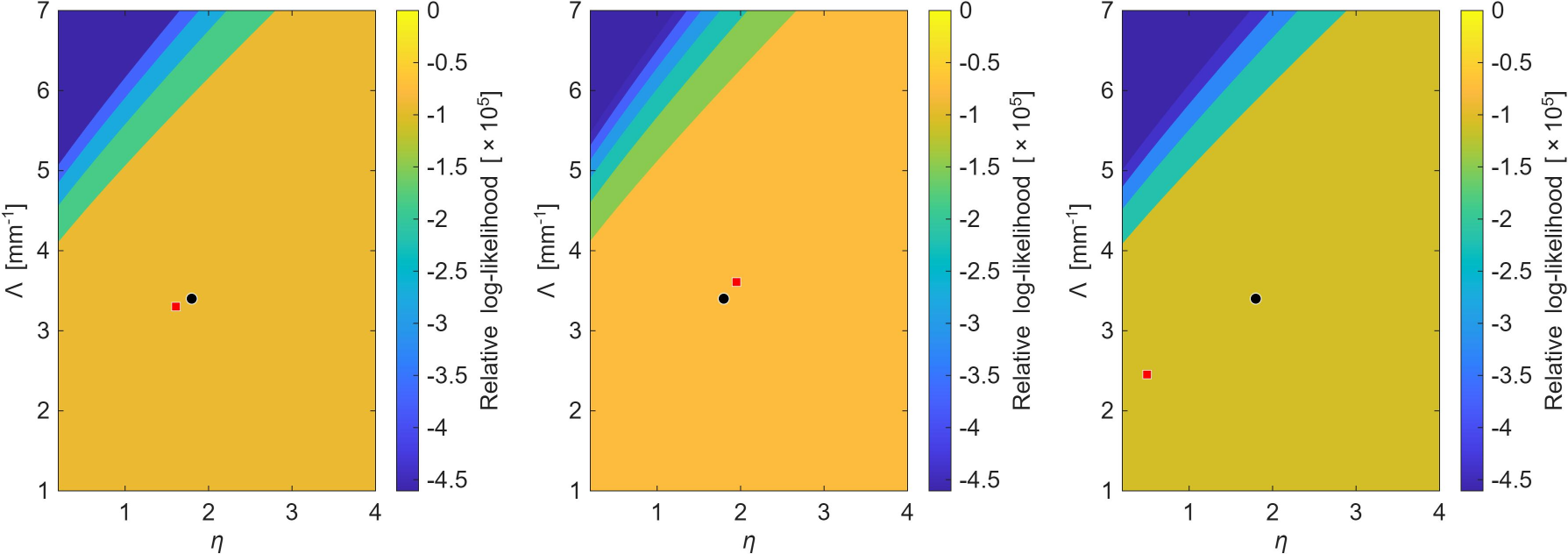}
\caption{Relative log-likelihood surfaces from the synthetic identifiability analysis. From left to right, the panels use the true wind parameters, $\sigma_v+0.2$~m/s with $\mu_v$ fixed, and $\mu_v+0.2$~m/s with $\sigma_v$ fixed. The filled circle marks the truth $(\eta,\Lambda)$, and the square marks the grid minimum of the plotted surface. The figure illustrates that the latent DSD pair is weakly identified even when the spectral fit remains good.}
\label{fig:identifiability}
\end{figure*}

Fig.~\ref{fig:identifiability} summarizes this behavior using the synthetic identifiability study. The surface is shown as a relative log-likelihood so that broad valleys and ridges remain visible rather than being compressed by a sharply peaked color scale. The figure is not meant to prove that the latent pair is unobservable in principle. Instead, it shows that the single-polarization short-CPI inverse problem is poorly conditioned in the $(\eta,\Lambda)$ coordinates, and that this poor conditioning explains why those parameters should not be emphasized as final retrieval products.

The practical consequence is that motion of the latent DSD pair along the broad likelihood ridge need not imply equally large changes in the derived quantities. In the present problem, $D_m$ and $V_T$ vary more slowly than $(\eta,\Lambda)$ along the weak-curvature direction, which is why they remain the more meaningful retrieval products for short-CPI single-polarization data.

\section{Simulation Design and Benchmarks}
\subsection{Simulated-Data Setup}
The simulated-data settings are chosen to reflect the radar conditions of interest while remaining within the simplified forward model. The real radar motivating the study is an X-band system operating at $f_c=\SI{9.4}{GHz}$, and the experiments considered here use the horizontal-polarization channel. For one polarization, the pulse repetition interval is $T=\SI{813.2}{\micro\second}$, which gives an unambiguous Doppler velocity of approximately \SI{9.8}{m/s}. The radar beamwidth is about $2.5^\circ$ in azimuth and $2^\circ$ in elevation, and the real-data experiment considered in the paper uses an elevation angle close to $30^\circ$ together with a fast azimuthal scan rate of $\Omega=5$ rpm. In the processed real-data product, each resolution cell contains 100 slow-time echoes, which are rearranged into short coherent records with $N=50$ and two incoherent PSD realizations per pixel. The simulated-data study mirrors that short-CPI regime by adopting the same carrier frequency, pulse repetition timing, elevation-angle dependence, CPI length, and incoherent aggregation logic. In all simulations $N_0$ is treated as known. This avoids conflating the main question of the paper with an additional amplitude-identification problem.

The benchmark is Chen-type retrieval from spectral moments \cite{Chen2020VerticalDistribution,Klaassen1989DeterminationRadar}. In that benchmark, the fall-spectrum mean and width are formed from the measured Doppler moments as $\hat{\bar{V}}_{T\psi}=\mu_r-\hat{\mu}^{(\mathrm{Chen})}_v$ and $\hat{\sigma}_{p\psi}^2=\sigma_r^2-\hat{\sigma}_{v}^{2(\mathrm{Chen})}$, after which the Chen inversion is applied to obtain $\hat{\Lambda}^{(\mathrm{Chen})}$ and $\hat{\eta}^{(\mathrm{Chen})}$. Because the Chen formulation does not estimate the wind parameters from the same PSD realization, the benchmark must be supplied with auxiliary wind information. In the present paper that auxiliary information is perturbed using short-CPI uncertainty models tied to $N$:
\begin{equation}
\hat{\mu}^{(\mathrm{Chen})}_v \sim \mathcal{N}\!\left(\mu_v^{(\mathrm{true})}, \left[\frac{2V_a}{N}\right]^2\right),
\end{equation}
\begin{equation}
\hat{\sigma}_{v}^{2(\mathrm{Chen})} \sim \mathrm{Gamma}\!\left(\frac{N-1}{2},\; \frac{2\sigma_{v,\mathrm{true}}^2}{N-1}\right),
\end{equation}
where $V_a$ is the unambiguous velocity. For the present short-CPI setting with $N=50$ and $V_a\approx \SI{9.8}{m/s}$, this corresponds to a standard deviation of about \SI{0.39}{m/s} for $\hat{\mu}^{(\mathrm{Chen})}_v$ together with an $N$-consistent dispersion in $\hat{\sigma}_v^2$. The benchmark should therefore be interpreted as a Chen-type retrieval supplied with imperfect external wind information, rather than as a moment method operating with exact wind knowledge.

This choice of benchmark is deliberate. The goal is not to compete with dual-polarization retrievals, but to compare two single-polarization strategies under short-CPI conditions: a full-spectrum likelihood approach and a reduced-moment approach. In that comparison, SPLIT is assessed where it is expected to be strongest, namely when several noisy incoherent spectra are available and when the detailed PSD shape contains useful information beyond low-order moments.

\subsection{Sensitivity to Wind Broadening}
\begin{figure*}[!t]
\centering
\begin{subfigure}[b]{0.48\linewidth}
    \centering
    \includegraphics[width=\linewidth,trim=0.08cm 0.08cm 0.08cm 0.08cm,clip]{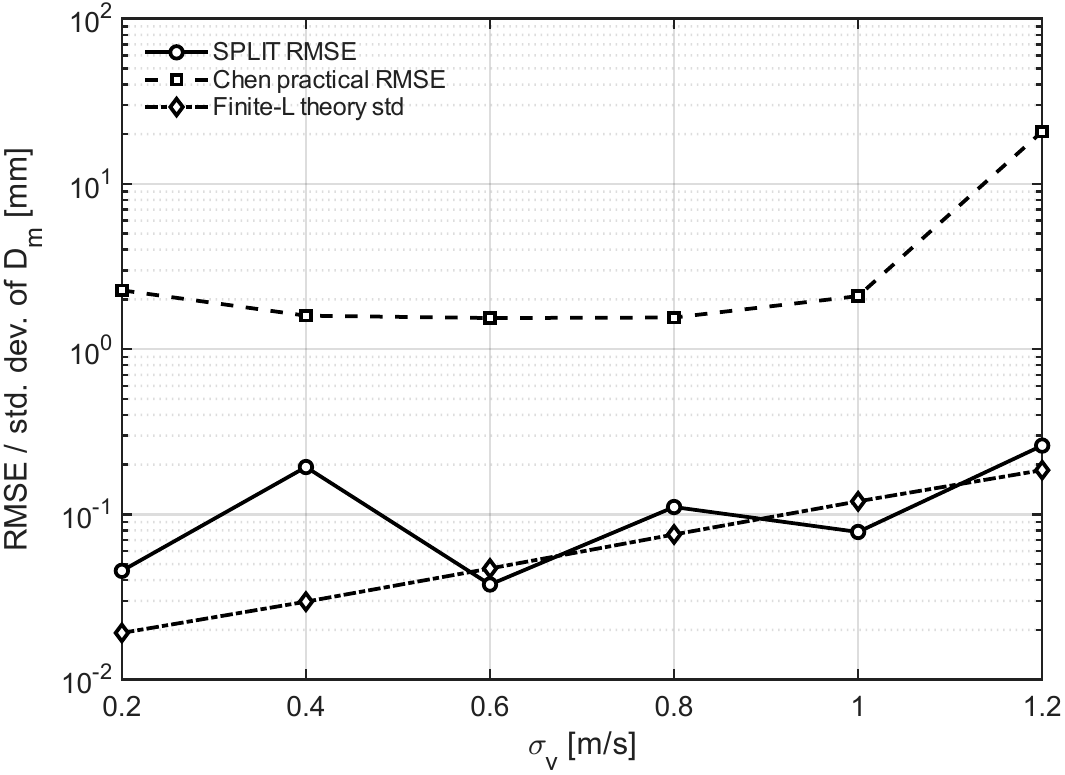}
    \caption{$D_m$ RMSE versus wind broadening.}
\end{subfigure}
\hfill
\begin{subfigure}[b]{0.48\linewidth}
    \centering
    \includegraphics[width=\linewidth,trim=0.08cm 0.08cm 0.08cm 0.08cm,clip]{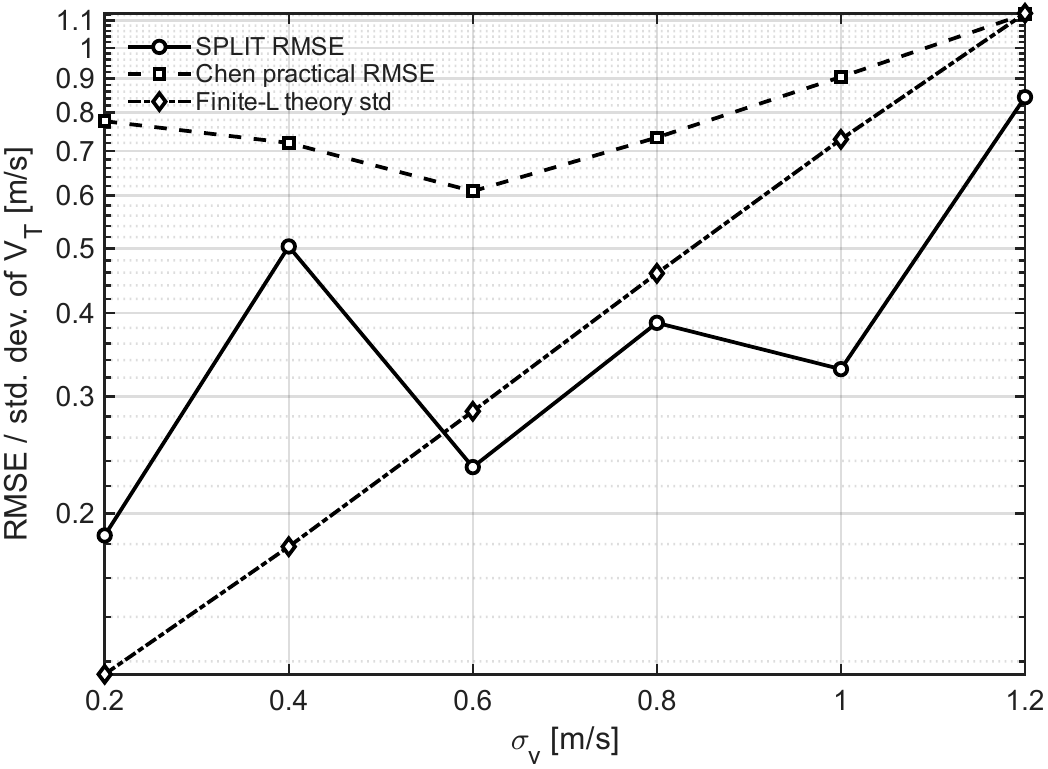}
    \caption{$V_T$ RMSE versus wind broadening.}
\end{subfigure}
\caption{Practical Chen benchmark versus SPLIT under increasing wind spectral width, shown on semi-logarithmic vertical axes. The SPLIT and Chen curves show empirical RMSE, while the third curve shows the finite-$L$ theoretical standard deviation propagated from the Fisher covariance of the latent parameter vector. In this sweep the fixed parameters are $\theta=29^\circ$, $\mu_v=5\cos 29^\circ \approx 4.37$~m/s, $\eta=1.8$, $\Lambda=3.4$, $N_0$ known, $N=50$, and $L=64$, while $\sigma_v$ is varied. SPLIT is most consistently advantageous for $D_m$, while the $V_T$ comparison is more conditional.}
\label{fig:sigma_sweep}
\end{figure*}

Fig.~\ref{fig:sigma_sweep} compares the error trends versus wind spectral width. The most robust finding is the strong improvement of SPLIT in $D_m$ relative to the Chen benchmark. The $V_T$ comparison is more nuanced and is therefore discussed more cautiously. The method is emphasized where it is genuinely advantageous, while the remaining limitations are interpreted through the identifiability analysis.

What matters here is whether the method retrieves the reported quantities accurately under realistic short-CPI conditions. The latent parameters remain part of the inverse model, but the simulation results are interpreted through $D_m$ and $V_T$ rather than through standalone latent-parameter sweeps.

\subsection{Benefit of Incoherent Aggregation}
\begin{figure*}[!t]
\centering
\begin{subfigure}[b]{0.48\linewidth}
    \centering
    \includegraphics[width=\linewidth,trim=0.08cm 0.08cm 0.08cm 0.08cm,clip]{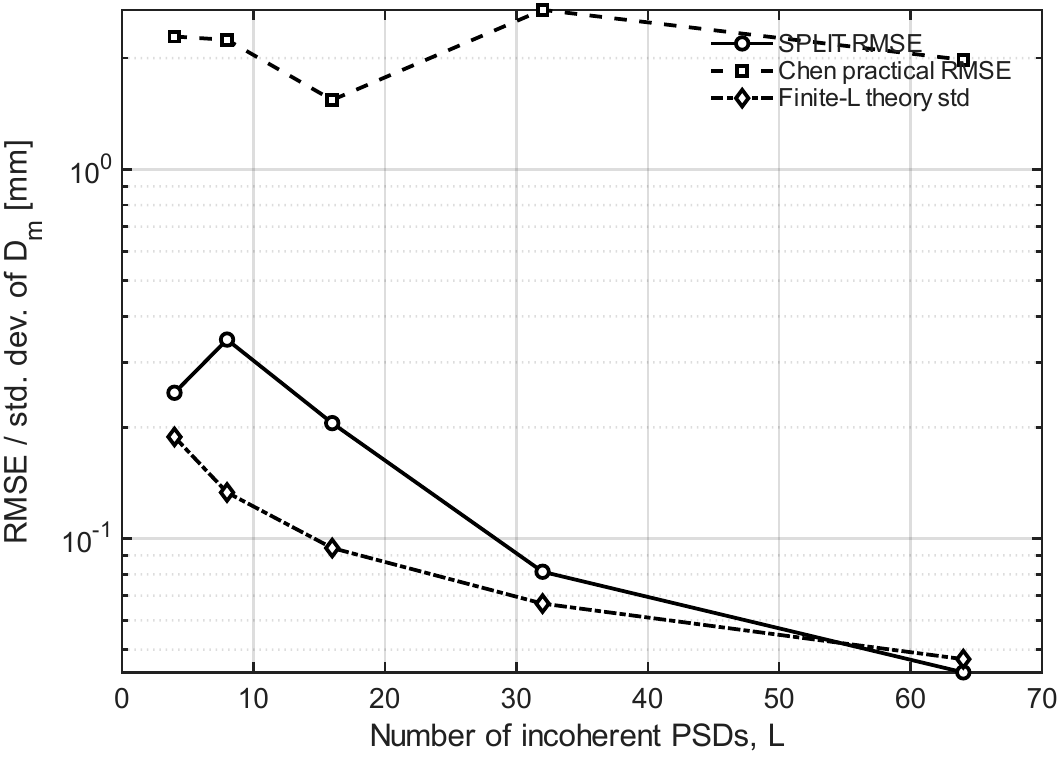}
    \caption{$D_m$ versus number of incoherent spectra.}
\end{subfigure}
\hfill
\begin{subfigure}[b]{0.48\linewidth}
    \centering
    \includegraphics[width=\linewidth,trim=0.08cm 0.08cm 0.08cm 0.08cm,clip]{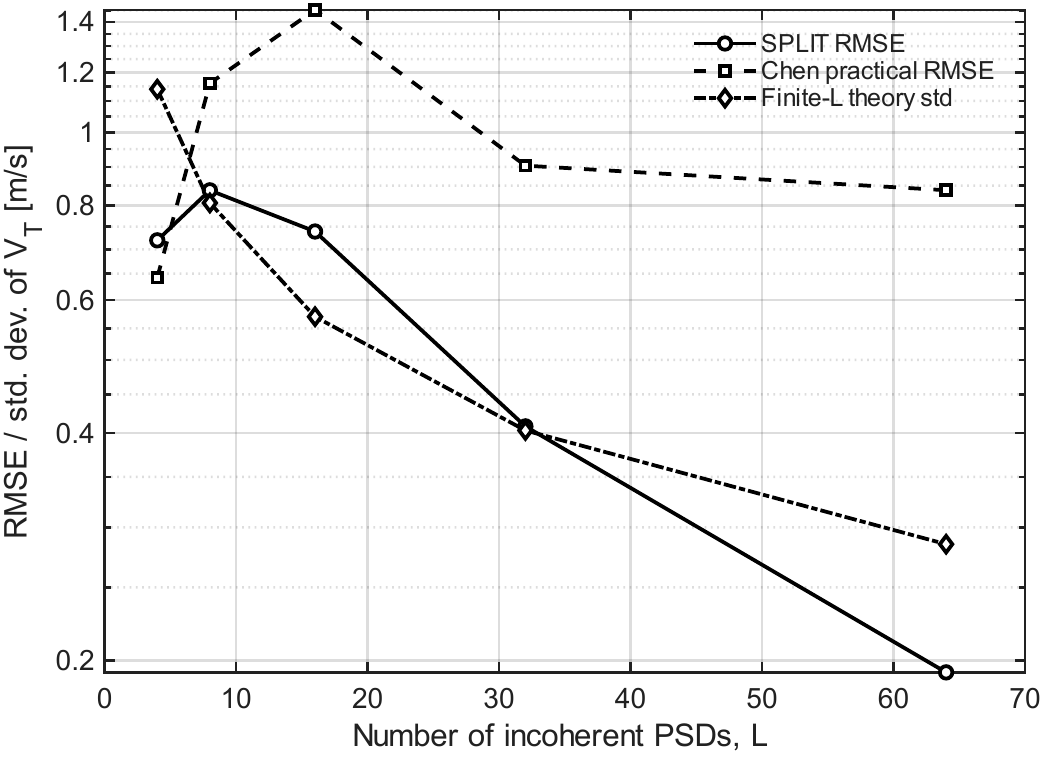}
    \caption{$V_T$ versus number of incoherent spectra.}
\end{subfigure}
\caption{Effect of the number of incoherent PSDs, shown on semi-logarithmic vertical axes, with SPLIT and Chen empirical errors plotted alongside the finite-$L$ theoretical standard deviation of the SPLIT-derived quantities. In this sweep the fixed parameters are $\theta=29^\circ$, $\mu_v=5\cos 29^\circ \approx 4.37$~m/s, $\sigma_v=0.8$~m/s, $\eta=1.8$, $\Lambda=3.4$, $N_0$ known, and $N=50$, while $L$ is varied.}
\label{fig:L_sweep}
\end{figure*}

Fig.~\ref{fig:L_sweep} shows the effect of increasing the number of incoherent PSDs. The empirical SPLIT RMSE is plotted together with a finite-$L$ theoretical standard deviation obtained from the Fisher covariance of the SPLIT parameter vector and propagated to $D_m$ and $V_T$. These overlays are intentionally presented as finite-record theoretical variance curves rather than as main-text CRB claims for the latent parameters. They are useful here because they provide a compact interpretation of the \mbox{$L$-sweep} without reintroducing unfavorable latent-parameter plots.

The role of Fig.~\ref{fig:L_sweep} is twofold. First, it supports the practical message that SPLIT benefits systematically from incoherent aggregation, which is precisely the operating mode of interest for fast-scanning radars. Second, it connects the empirical trend to a theory-based finite-$L$ uncertainty proxy without overselling the asymptotic CRB as a full explanation of finite-sample behavior.

\subsection{Effect of Elevation Angle}
\begin{figure*}[!t]
\centering
\begin{subfigure}[b]{0.48\linewidth}
    \centering
    \includegraphics[width=\linewidth,trim=0.08cm 0.08cm 0.08cm 0.08cm,clip]{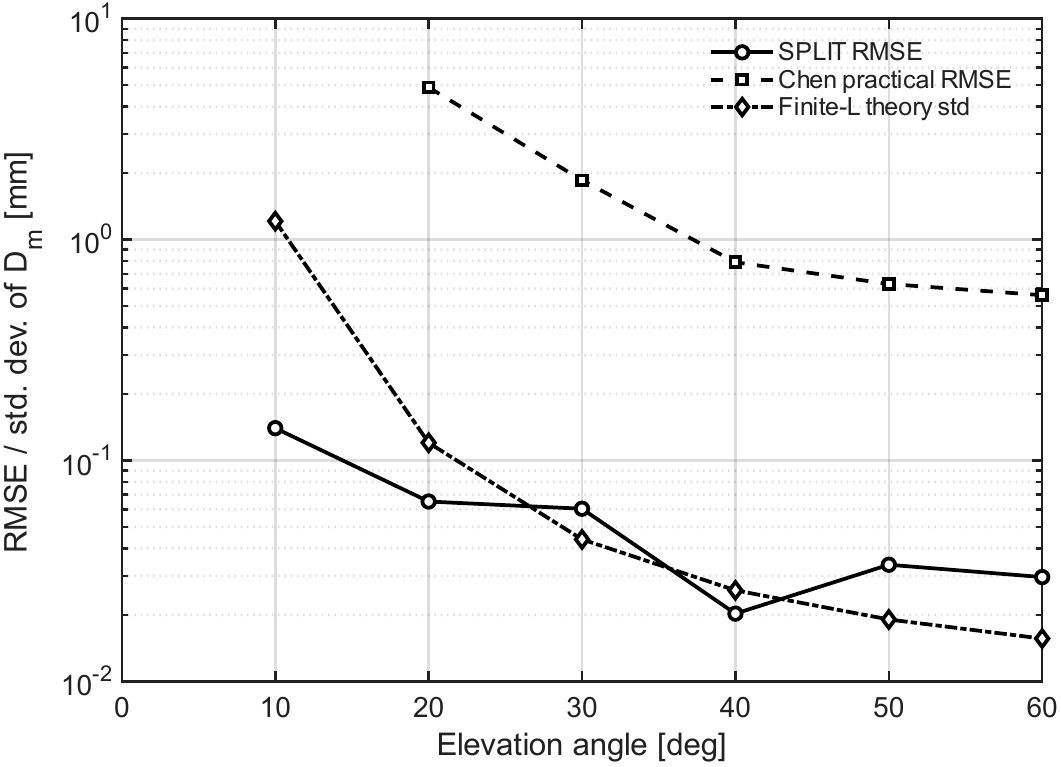}
    \caption{$D_m$ RMSE versus elevation.}
\end{subfigure}
\hfill
\begin{subfigure}[b]{0.48\linewidth}
    \centering
    \includegraphics[width=\linewidth,trim=0.08cm 0.08cm 0.08cm 0.08cm,clip]{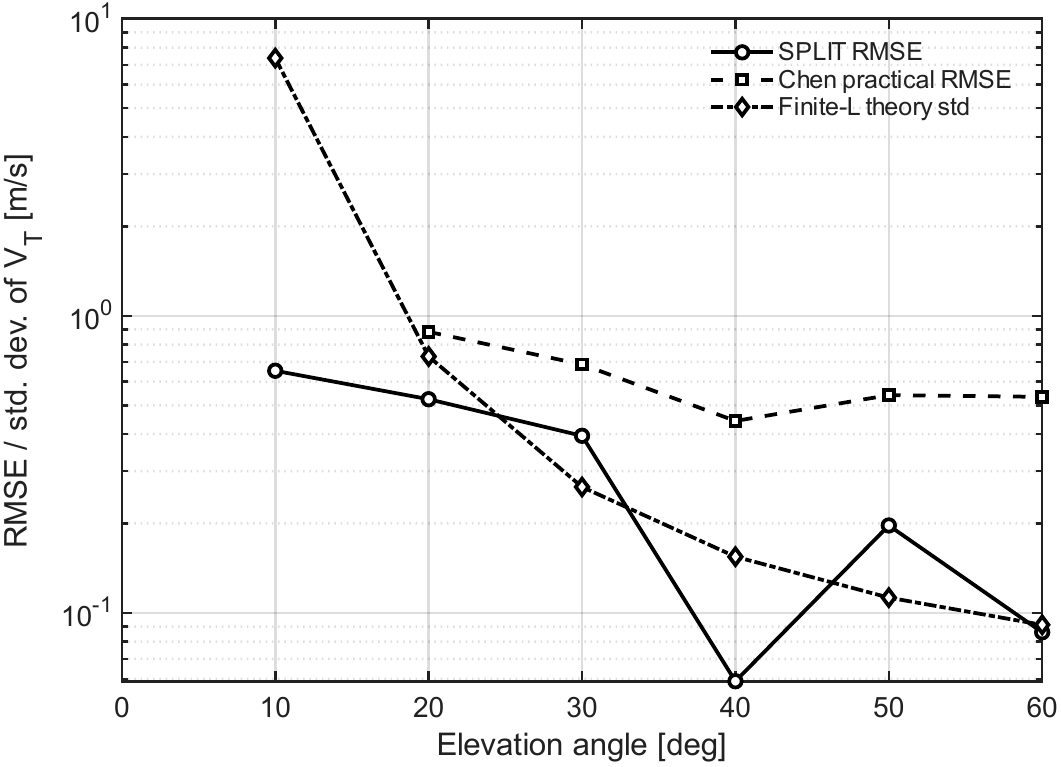}
    \caption{$V_T$ RMSE versus elevation.}
\end{subfigure}
\caption{Elevation-angle study for the practical benchmark, shown on semi-logarithmic vertical axes. The SPLIT and Chen curves show empirical RMSE, while the third curve shows the finite-$L$ theoretical standard deviation propagated from the Fisher covariance of the latent parameter vector. In this sweep the fixed parameters are $\sigma_v=0.8$~m/s, $\eta=1.8$, $\Lambda=3.4$, $N_0$ known, $N=50$, and $L=64$; the radial-wind mean follows the geometry as $\mu_v=5\cos\theta$. The Chen 10$^\circ$ point is undefined under the moment inversion used here, which is why its visible curve begins at 20$^\circ$; the 0$^\circ$ case is not shown because the present fall-speed projection model becomes singular as $\sin\theta \rightarrow 0$. The higher-elevation cases are better conditioned for the derived quantities, especially when compared with the short-CPI single-polarization alternatives.}
\label{fig:elevation_sweep}
\end{figure*}

The elevation-angle dependence remains important. Higher elevation changes the projection between terminal fall speed and radial velocity, which affects both sensitivity and parameter coupling. Fig.~\ref{fig:elevation_sweep} therefore supports the interpretation of when the single-polarization likelihood is best conditioned.

The elevation study also helps interpret the real-data experiment. It provides a controlled theoretical explanation for why certain scan angles offer a more favorable tradeoff between wind coupling and terminal-velocity sensitivity. This is valuable because the real radar observations are taken at a specific operational elevation, whereas the simulations are used to map the broader conditioning landscape. At very low elevation, the Chen moment inversion becomes ill-conditioned, and the present projected fall-speed model is singular as $\theta \rightarrow 0^\circ$ because the fall contribution vanishes with $\sin\theta$. The low-angle behavior in Fig.~\ref{fig:elevation_sweep} should therefore be read as a conditioning limit of the model-based decomposition, not as a universal statement about scanning radars in general.

\section{Real-Data Experiment: MESEWI, Green Village, and Zweth}

\subsection{Real-Data Processing and Same-Scan Spatial Likelihood Pooling}
\begin{figure*}[!t]
\centering
\includegraphics[width=\linewidth,trim=0.05cm 0.05cm 0.05cm 0.05cm,clip]{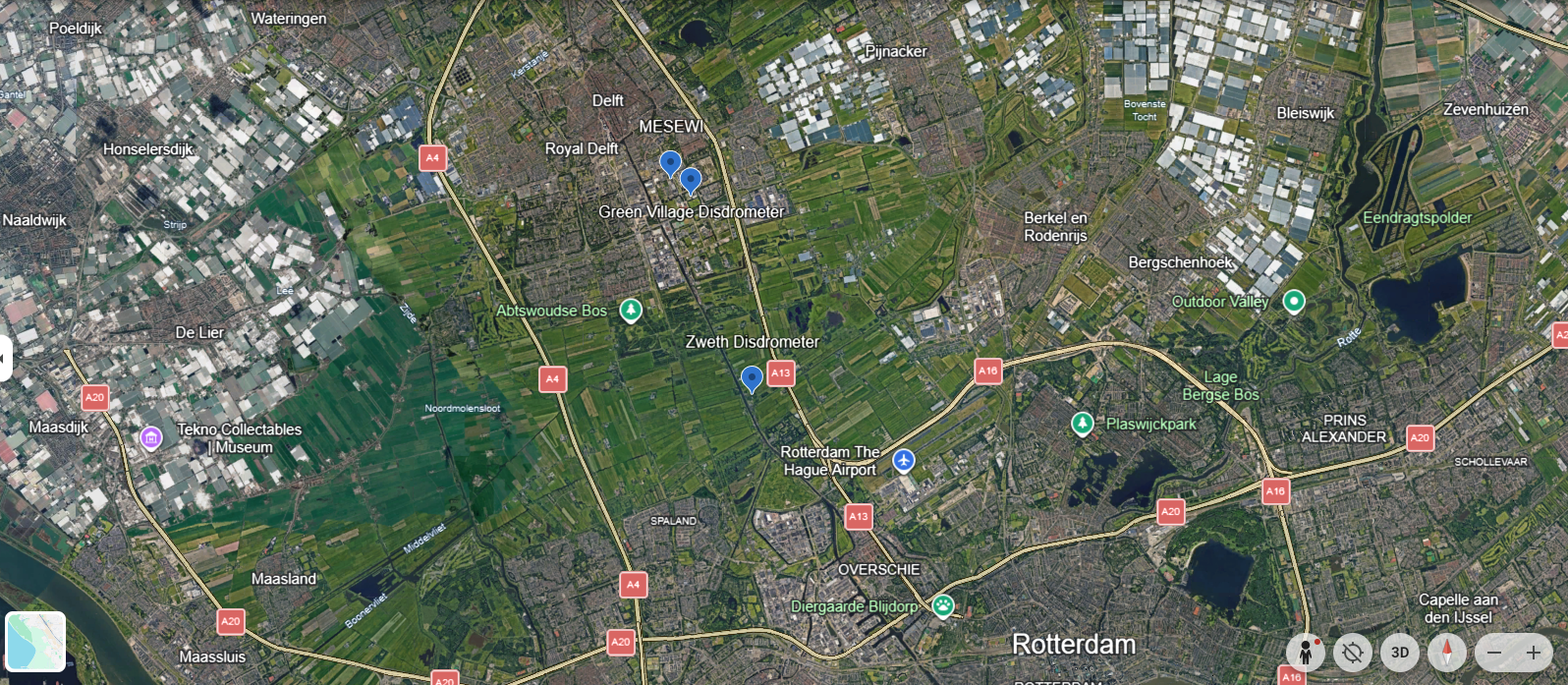}
\caption{Site map for the real-data experiments. The regional panel shows the MESEWI radar, Green Village, and Zweth, and the inset enlarges the campus-scale geometry of the MESEWI radar and Green Village. The map is shown on a satellite basemap for geographic context. The MESEWI marker corresponds to the radar position inferred from the trusted Green Village collocation used for the radar--disdrometer comparison.}
\label{fig:site_map}
\end{figure*}

The real-data analysis uses the MESEWI fast-scanning X-band radar data acquired on 9 May 2023. The MESEWI system is a fully polarimetric \SI{9.4}{GHz} radar operated at TU Delft; the present study uses the HH channel only. The relevant configuration for the experiment is a pulse repetition interval of \SI{813.2}{\micro\second}, an unambiguous Doppler velocity of about \SI{9.8}{m/s}, beamwidths of approximately $2.5^\circ$ in azimuth and $2^\circ$ in elevation, a working elevation near $30^\circ$, and an azimuthal scan rate of $\Omega=5$ rpm. Fig.~\ref{fig:site_map} summarizes the site geometry. The MESEWI radar is located on the TU Delft campus, Green Village is in the immediate near range of the radar, and Zweth is several kilometers farther south. This geometry allows the same retrieval formulation to be assessed both in the near range and at a substantially larger slant range.

For each scan, the calibrated radar time series are first converted into range-resolved Doppler spectra by range processing the complex returns, removing the slow-time mean, and estimating the Doppler PSD on short coherent records. Each resolution cell contains 100 slow-time echoes for Doppler processing. Those echoes are rearranged into two incoherent PSD realizations of length $N=50$, together with a pulse-pair mean Doppler estimate and a reflectivity proxy. These per-pixel spectra are then used as the building blocks of the same-scan retrieval.

The key change relative to a pointwise retrieval is that the incoherent sample support is increased spatially rather than temporally. For an anchor pixel $p$, let $\mathcal{S}_p$ denote a set of neighboring pixels drawn from the same scan, selected subject to similarity constraints in pulse-pair mean Doppler velocity and reflectivity. Assuming that the latent vector $\bm{\theta}=[\mu_v,\sigma_v,\eta,\Lambda]^\top$ is locally homogeneous across that neighborhood, the retrieval is formed by pooling the PSDs directly in the likelihood:
\begin{equation}
\hat{\bm{\theta}}_p
=
\arg\max_{\bm{\theta}}
\sum_{q \in \mathcal{S}_p}\sum_{\ell=1}^{L_{\mathrm{pers}}}
\log p\!\left(Z_{q,\ell}\mid \bm{\theta}\right).
\label{eq:local_spatial_pool}
\end{equation}
Here $L_{\mathrm{pers}}=2$ is the number of processed incoherent PSDs available per pixel in one scan, and the effective number of spectra is therefore
\begin{equation}
L_p = |\mathcal{S}_p|L_{\mathrm{pers}}.
\end{equation}
For the Green Village and Zweth time-series comparisons, the local support is fixed to 31 pixels, giving $L_p=62$ for each scan. For the PPI product, the same likelihood-pooling principle is applied anchor by anchor, with the resulting local estimates averaged back onto the display grid where the local supports overlap.

The adaptive near-zero-bin removal is applied after the local ensemble has been formed. That is, the notch width is chosen from the pooled local spectrum and then the likelihood is evaluated on the retained Doppler bins. In this way, the short-CPI spectra remain the primitive observations, and aggregation occurs at the likelihood level rather than by averaging separately retrieved parameter estimates.

In the real-data examples, the intercept parameter is held fixed using a site-specific concentration proxy from the colocated disdrometer over the 07:46--07:53~UTC comparison window. For Green Village, the adopted value is $N_0=3829$; for Zweth, the corresponding window-integrated proxy obtained from the corrected Parsivel \texttt{raw\_data} matrix is $N_0=3882$. The two values are close, which is helpful for interpreting differences between the two sites as primarily geometric and spectral rather than simply amplitude-driven. A light-rain attenuation proxy is also retained in the reflectivity model so that the reflectivity term does not artificially force the wind-width parameter to absorb amplitude mismatch.

\subsection{Same-Scan PPI Retrieval}
\begin{figure*}[!t]
\centering
    \includegraphics[width=\linewidth]{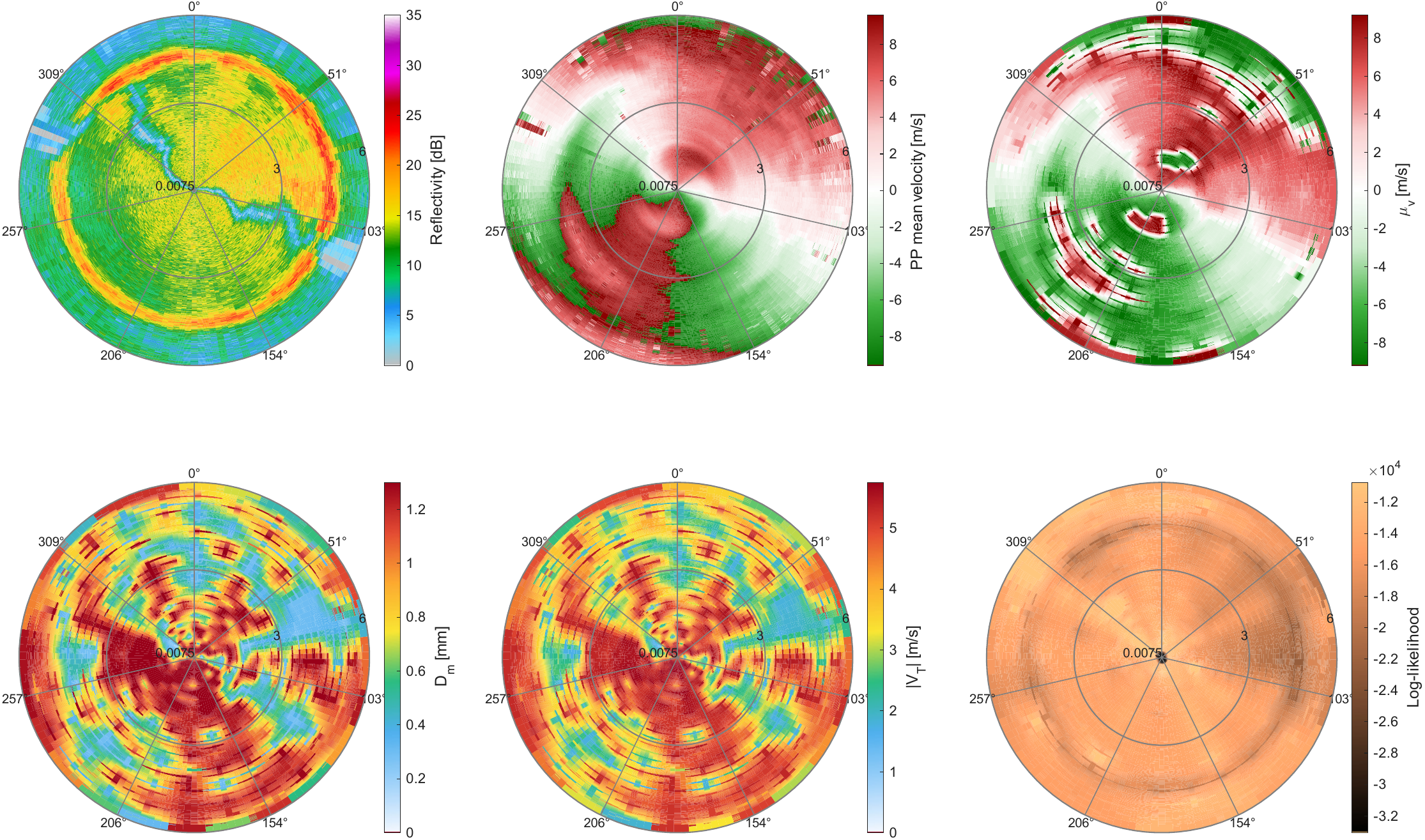}
\caption{Same-scan PPI retrieval over the first 800 range bins. The top row shows the measured reflectivity proxy, the measured pulse-pair mean Doppler velocity, and the SPLIT estimate of the radial-wind mean $\mu_v$. The bottom row shows the SPLIT estimates of $D_m$ and $|V_T|$, together with the optimized local log-likelihood field. Each local estimate is obtained from a pooled same-scan neighborhood, and overlapping local supports are averaged back onto the PPI. The log-likelihood panel uses a clipped color scale so that lower-likelihood sectors remain visible rather than being compressed by the highest-power rain regions.}
\label{fig:ppi_spatial_same_scan}
\end{figure*}

The same-scan spatial strategy can be extended from one neighborhood to a broader plan-position-indicator (PPI) product. Fig.~\ref{fig:ppi_spatial_same_scan} shows one such retrieval for the scan containing the Green Village case. The figure is displayed with the same north-referenced polar convention used in the earlier PSE presentation of the MESEWI data, so that $\phi=0^\circ$ corresponds to geographic north and azimuth increases clockwise. Only the first 800 range bins are shown, which keeps the display below the broader melting-layer regime and concentrates the analysis on the rain-dominated portion of the scan for which the single-rain forward model is intended.

The top row of Fig.~\ref{fig:ppi_spatial_same_scan} contains the measured reflectivity proxy, the measured pulse-pair mean Doppler velocity, and the SPLIT estimate of $\mu_v$. The bottom row contains the SPLIT estimates of $D_m$ and $|V_T|$, together with the optimized local log-likelihood field.

Several points are evident in Fig.~\ref{fig:ppi_spatial_same_scan}. First, the full scan can be mapped efficiently even though each individual Doppler spectrum is based on only a few slow-time samples. This is one of the main practical novelties of the method: the retrieval is not restricted to a single disdrometer-like point comparison, but yields spatially resolved fields of $\mu_v$, $D_m$, and $V_T$ from short-CPI data by pooling same-scan spectra locally in space. Second, the reflectivity field is smoother than the retrieved $D_m$ and $|V_T|$ fields, which is expected because reflectivity is closer to a direct observable while the latter two fields come from the inverse fit. Third, the broad azimuthal organization in $\mu_v$ is physically consistent with the beam-projected wind contribution, while the close correspondence between $D_m$ and $|V_T|$ reflects the fact that both are derived from the same latent $(\eta,\Lambda)$ pair. Finally, the low-reflectivity wedge-like sector persists even when only one scan is used and the aggregation is purely spatial. This indicates that the sector is not created by temporal stacking. Its coincidence with near-zero mean Doppler makes a preprocessing-sensitive clutter-overlap interpretation plausible: when the precipitation spectrum lies too close to zero Doppler, part of the rain contribution can be attenuated together with the clutter-suppression region.

\begin{figure*}[!t]
\centering
\includegraphics[width=\linewidth]{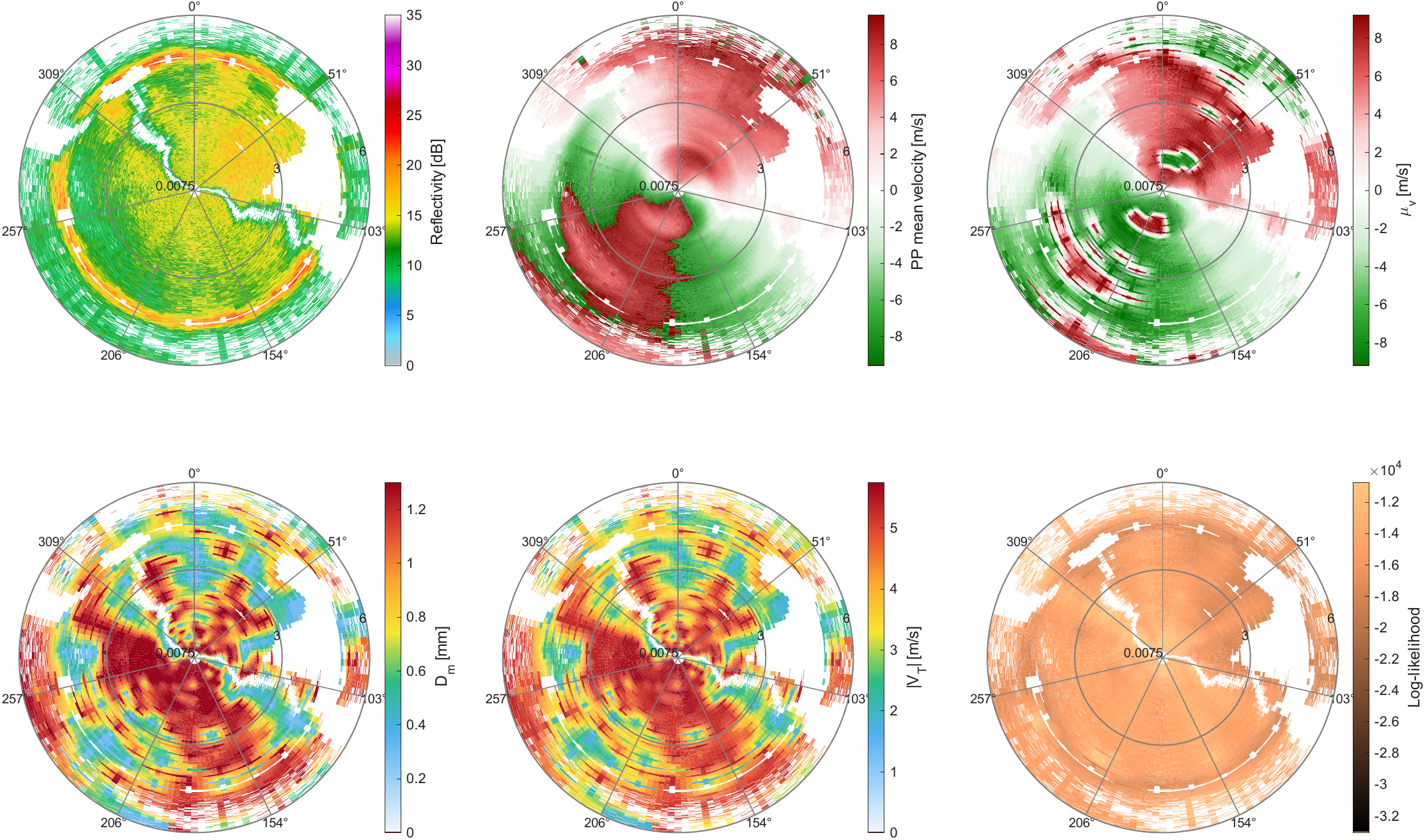}
\caption{Same-scan PPI retrieval after a relaxed joint quality mask based on minimum reflectivity and minimum optimized log-likelihood. The panel order is the same as in Fig.~\ref{fig:ppi_spatial_same_scan}, and rejected regions are shown in white. The masked view removes both weak-signal regions and bright-return regions whose spectra remain poorly matched by the rain-only SPLIT model. Thus, a sector can be rejected either because the return is too weak to constrain the fit or because the observed spectrum is energetic but spectrally incompatible with the single-rain model.}
\label{fig:ppi_spatial_same_scan_masked}
\end{figure*}

To emphasize the regions where the decomposition is most trustworthy, Fig.~\ref{fig:ppi_spatial_same_scan_masked} applies a joint quality mask using both reflectivity and optimized log-likelihood. The reflectivity threshold is taken as the larger of 8~dB and the lower decile of the displayed reflectivity values, while the likelihood threshold is taken from the lower decile of the local log-likelihood field. This mask is intentionally relaxed, so that part of the high-reflectivity sector between approximately $0^\circ$ and $100^\circ$ azimuth remains visible.

\begin{figure*}[!t]
\centering
\includegraphics[width=0.98\linewidth,trim=0.05cm 0.05cm 0.05cm 0.05cm,clip]{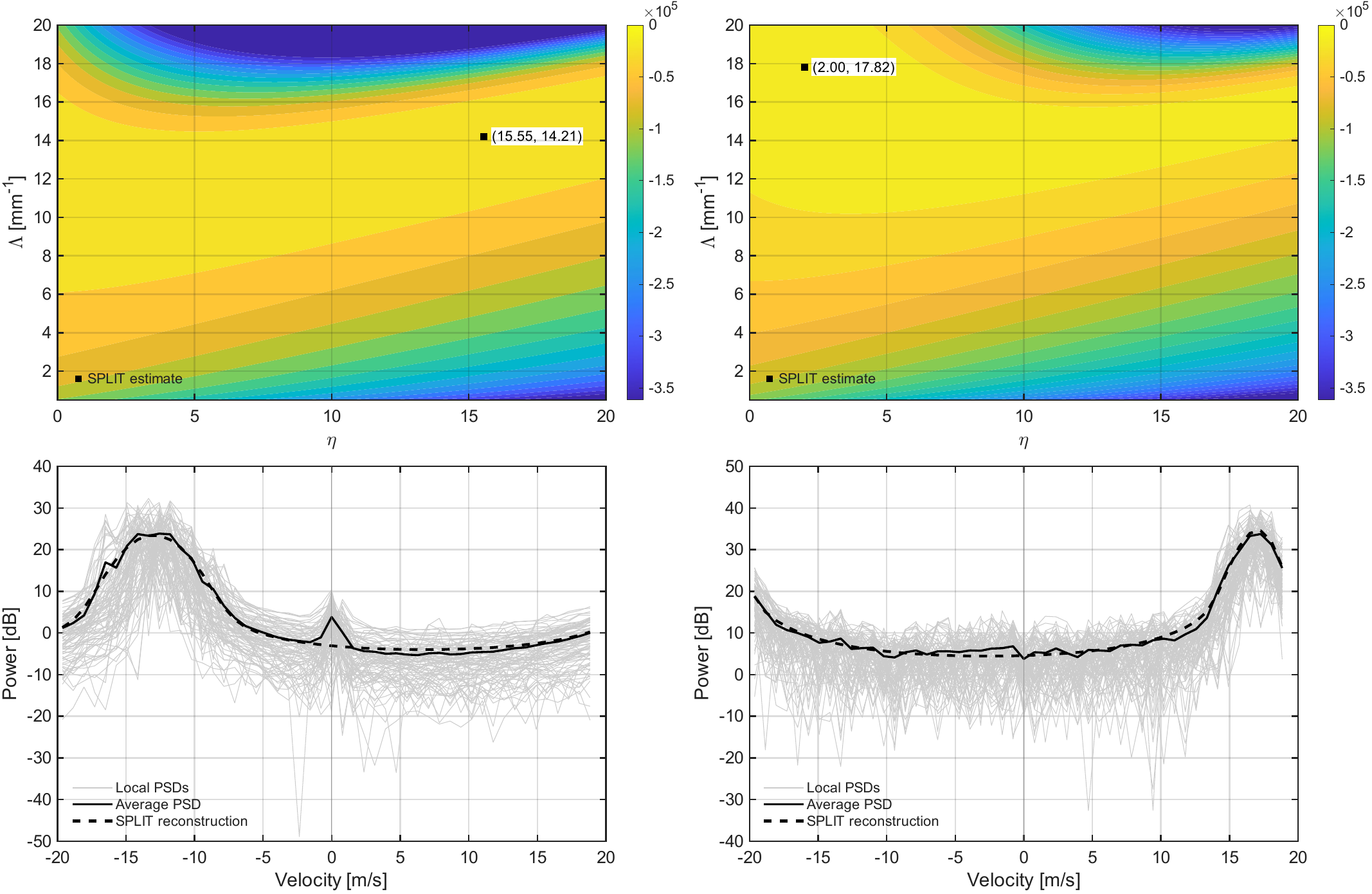}
\caption{Local fit-quality examples from the same scan used in the PPI retrieval. The left column shows a retained local aggregate, and the right column shows a rejected local aggregate. In the top row, $\mu_v$ and $\sigma_v$ are fixed at the local SPLIT estimate and the marker denotes the corresponding $(\eta,\Lambda)$ pair. In the bottom row, the gray curves are the pooled local PSD measurements, the solid black curve is their mean, and the dashed black curve is the fitted SPLIT spectrum. The figure illustrates why high reflectivity alone is not sufficient for a trustworthy retrieval: a sector can remain bright but still be rejected when its local spectral shape is poorly matched by the single-rain semianalytical model.}
\label{fig:fit_examples}
\end{figure*}

The meaning of the mask becomes clearer when the local fits are inspected directly. Fig.~\ref{fig:fit_examples} compares one retained local aggregate and one rejected local aggregate from the same scan. In both cases, the top row shows the local likelihood slice in the $(\eta,\Lambda)$ plane after fixing $\mu_v$ and $\sigma_v$ at their local SPLIT values, and the bottom row shows the corresponding local PSD ensemble in gray together with the mean measured spectrum and the fitted SPLIT reconstruction. The retained example exhibits both a broader high-likelihood basin and a visibly better agreement between the pooled measured spectra and the forward model. The rejected example, by contrast, is bright in reflectivity but remains less compatible with the rain-only model. This figure is therefore the local spectral counterpart of the masked PPI: it shows that the quality screen is not merely suppressing weak power, but is actively separating strong-yet-model-incompatible spectra from strong-and-well-explained spectra.

\subsection{Green Village and Zweth Time Series}
\begin{figure*}[!t]
\centering
\begin{subfigure}[t]{0.98\linewidth}
    \centering
    \includegraphics[width=\linewidth,trim=0.05cm 0.05cm 0.05cm 0.05cm,clip]{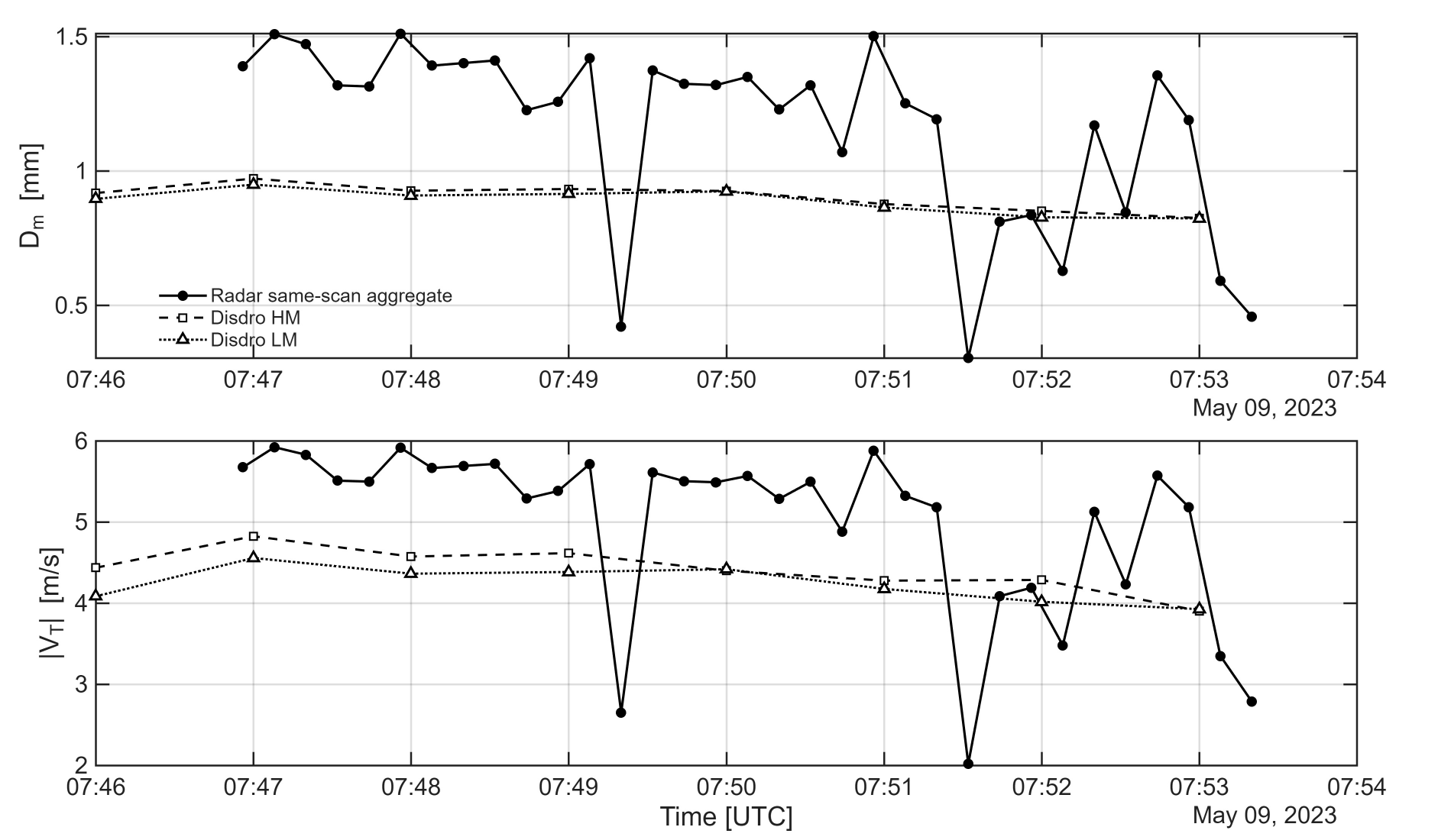}
    \caption{Green Village same-scan retrievals over time.}
\end{subfigure}

\medskip

\begin{subfigure}[t]{0.98\linewidth}
    \centering
    \includegraphics[width=\linewidth,trim=0.05cm 0.05cm 0.05cm 0.05cm,clip]{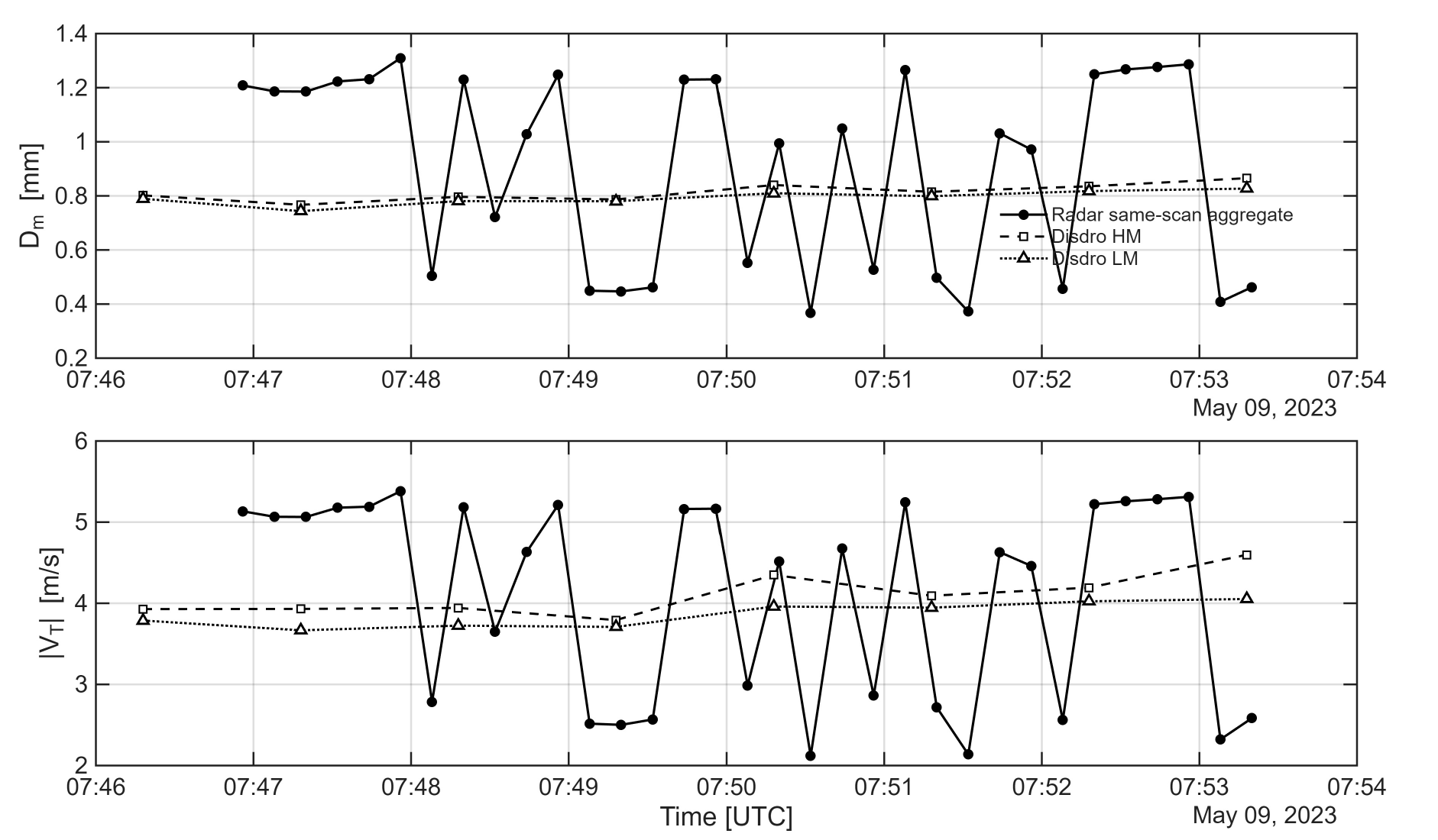}
    \caption{Zweth same-scan retrievals over time.}
\end{subfigure}
\caption{Time evolution of the same-scan spatially aggregated radar retrievals at the two disdrometer sites. In each panel the black solid curve shows the per-scan radar retrieval, while the dashed-square and dotted-triangle curves show the higher-moment and lower-moment disdrometer products, respectively. Both sites use $L=62$ same-scan spectra per radar estimate, obtained from 31 neighboring pixels with two incoherent PSDs per pixel.}
\label{fig:site_timeseries}
\end{figure*}

The same-scan spatial aggregate can also be evaluated scan by scan over the event. The MESEWI radar scanned at $\Omega=5$~rpm, corresponding to approximately 12~s between successive azimuth sweeps. Fig.~\ref{fig:site_timeseries} compares one same-scan spatial SPLIT estimate per scan with the one-minute disdrometer products over 07:46--07:53~UTC for both Green Village and Zweth. In both cases the radar estimate is formed from a 31-pixel same-scan neighborhood, so that the effective number of incoherent spectra is $L=62$.

The Green Village comparison is valuable because the temporal evolution is tracked well despite a non-negligible absolute bias. The radar retrieval follows the broad decrease in both $D_m$ and $|V_T|$ through the event, but remains systematically high on average. Over the full 33-scan window, the radar means are $\overline{D_m}=1.16$~mm and $\overline{|V_T|}=4.99$~m/s, compared with disdrometer higher-moment means of 0.90~mm and 4.42~m/s. This is consistent with the interpretation already suggested by the PPI fields: the method captures the event evolution, while the remaining bias reflects the combination of volume-versus-point mismatch, fall-time and advection differences between beam height and surface, and the residual sensitivity of the latent shape--slope pair to preprocessing and local heterogeneity.

The Zweth comparison is encouraging in a different way. The Zweth mean retrieval is closer to the disdrometer on average, particularly for $|V_T|$. Over the same 33-scan window, the radar means are $\overline{D_m}=0.91$~mm and $\overline{|V_T|}=4.10$~m/s, while the Zweth higher-moment disdrometer means are 0.81~mm and 4.10~m/s. At the same time, the radar retrieval varies more strongly from scan to scan at Zweth than at Green Village. This is physically reasonable because Zweth lies at a much larger range from the radar. The larger range implies a larger effective radar sampling volume, a lower signal-to-noise ratio, and stronger sensitivity to unresolved spatial heterogeneity inside the local aggregate. Thus, the Zweth time series demonstrates both a good mean agreement and a larger variance, which is entirely plausible for a more distant target.

\section{Discussion}
Three points emerge clearly from the analysis.

First, SPLIT is not a dual-polarization replacement. Its value is that it extracts useful microphysical proxies from single-polarization Doppler spectra when the data are limited to short incoherent records. It also differs from Doppler-spectrum inversion studies designed around profiler-style or long-observation settings \cite{Unal2015High-resolutionRadar,Wakasugi1986ASpectra,Rajopadhyaya1993MeasuringRadar}.

Second, the latent DSD pair is not the right place to claim strength. The log-likelihood analyses show that $(\eta,\Lambda)$ is weakly identified. The present presentation therefore avoids latent-parameter maps, large latent-parameter sweeps, and any wording suggesting that SPLIT provides a generally robust DSD-parameter product under the present measurement constraints.

Third, the real-data comparisons show two complementary strengths. At Green Village, the radar retrieval reproduces the event evolution well and therefore demonstrates that the method can monitor local temporal changes with only short-CPI same-scan data. At Zweth, the average radar retrieval is noticeably closer to the disdrometer, especially for $|V_T|$, even though the scan-to-scan variance is larger because the range is much greater. Taken together, these two sites show that the method is informative both as a spatial mapper and as a local time-series retrieval, while also illustrating how the bias-variance balance changes with range.

Fourth, the same-scan PPI study clarifies the role of preprocessing-sensitive sectors. The depressed-reflectivity wedge persists even in the strict single-scan product and is therefore not a by-product of temporal aggregation in the SPLIT stage. Its coincidence with near-zero mean Doppler makes a clutter-overlap interpretation plausible: when the precipitation spectrum lies too close to the clutter-suppression region, the preprocessing can attenuate part of the rain energy together with the unwanted component. Such sectors are therefore treated conservatively through the joint reflectivity-likelihood mask rather than through a more elaborate clutter submodel at this stage.

The discussion also clarifies the role of the latent parameters in the present framework. They are not discarded because they remain essential to the spectrum model. Rather, they are demoted from reported products to internal variables of the retrieval. This makes the paper more honest about what single-polarization short-CPI data can and cannot support, while retaining the practical advantage that spatially resolved $D_m$ and $V_T$ fields can still be formed efficiently from very short Doppler records.

\section{Conclusion}
This paper reformulates the single-polarization precipitation-retrieval problem for fast-scanning X-band weather radar. The same forward model is retained, but the interpretation is changed. Gamma-DSD parameters are treated as latent variables used to fit the Doppler spectrum, while radial-wind parameters, terminal fall velocity, and median diameter are treated as the reported outputs.

Four conclusions emerge. First, the likelihood surfaces and the local Fisher-curvature argument show directly why the latent DSD parameters are poorly conditioned. Second, the simulation study demonstrates where SPLIT is genuinely advantageous, particularly for $D_m$ under short-CPI incoherent operation. Third, the real-data study shows that same-scan spatial likelihood pooling can produce not only pointwise retrievals but also spatially resolved PPI fields of $D_m$ and $V_T$ from very short Doppler records. Fourth, the Green Village and Zweth comparisons show that useful temporal behavior and plausible local averages can be obtained from the same framework, even though site-dependent biases and range-dependent variance remain.

In that sense, SPLIT is best presented as a complementary single-polarization method for fast-scanning radars that cannot exploit a separable clear-air component and do not rely on dual-polarization microphysical retrieval. The same-scan PPI experiments further show that preprocessing-sensitive near-zero-Doppler sectors can be identified but are not yet fully recoverable without more explicit clutter-aware modeling, which is left for future work.

\appendix

\section{Approximation Validity and Decay Study}
SPLIT is derived from a finite-$N$ PSD approximation whose validity depends on how quickly the precipitation covariance kernel decays with lag. The purpose of this appendix is to make that statement explicit. If the relevant lag-domain support of the precipitation term is short compared with the available coherent record length, then the finite-$N$ approximation captures the part of the covariance that materially affects the Doppler spectrum. If the decay is too slow, then a larger coherent record or a higher-order quadrature treatment would be needed.

Let $G_q(\eta,\Lambda)$ denote the precipitation covariance term in the lag domain. The decay study is based on two simple diagnostics:
\begin{equation}
q_{1/2}:\ |G_q| = \tfrac{1}{2}|G_0|,
\qquad
q_{e}:\ |G_q| = e^{-1}|G_0|,
\end{equation}
which are referred to here as the half-life and the $e^{-1}$ lifetime of the precipitation kernel. These lag indices summarize how many coherent samples are needed to cover the substantially correlated part of the fall-velocity contribution.

Because $G_q$ does not admit a convenient closed form as a function of $q$, the numerical reference is obtained directly from the integral representation of the precipitation term. To interpret that decay more simply, an analytical proxy can be built from the approximate fall-spectrum width used in \cite{Chen2020VerticalDistribution}. Let $\sigma_W$ denote the corresponding reflectivity-weighted fall-spectrum width and define the normalized width
\begin{equation}
\sigma_{fWn}=\frac{\sigma_W}{2V_a},
\end{equation}
where $V_a$ is the unambiguous velocity. If one maps this width to a Gaussian-like decay proxy, then the associated half-life and lifetime are
\begin{equation}
q_{1/2}^{(\mathrm{a})}=
\frac{\sqrt{\log 2}}{\sqrt{2\pi^2}\,\sigma_{fWn}},
\qquad
q_{e}^{(\mathrm{a})}=
\frac{1}{\sqrt{2\pi^2}\,\sigma_{fWn}}.
\end{equation}
These expressions are not used to claim that $G_q$ is Gaussian. They are used only as a convenient analytical scale against which the numerical decay of the true precipitation kernel can be compared.

\begin{figure*}[!t]
\centering
\begin{subfigure}[b]{0.48\linewidth}
    \centering
    \includegraphics[width=\linewidth,trim=0.08cm 0.08cm 0.08cm 0.08cm,clip]{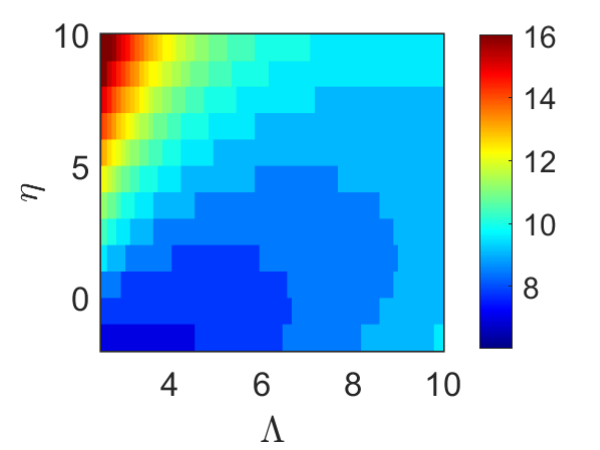}
    \caption{Numerical lifetime $q_e$ from the integral form of $G_q$.}
\end{subfigure}
\hfill
\begin{subfigure}[b]{0.48\linewidth}
    \centering
    \includegraphics[width=\linewidth,trim=0.08cm 0.08cm 0.08cm 0.08cm,clip]{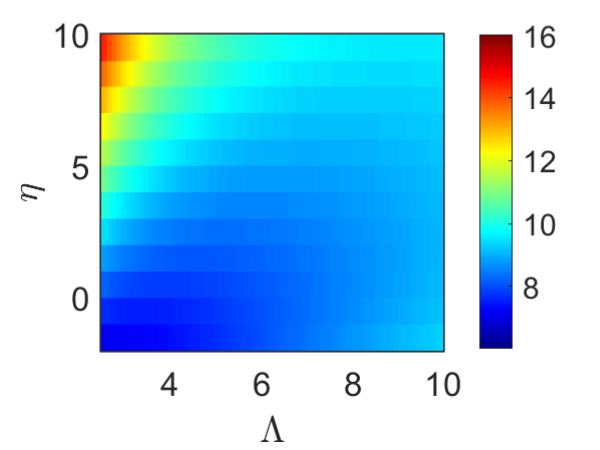}
    \caption{Analytical lifetime proxy based on $\sigma_{fWn}$.}
\end{subfigure}

\begin{subfigure}[b]{0.48\linewidth}
    \centering
    \includegraphics[width=\linewidth,trim=0.08cm 0.08cm 0.08cm 0.08cm,clip]{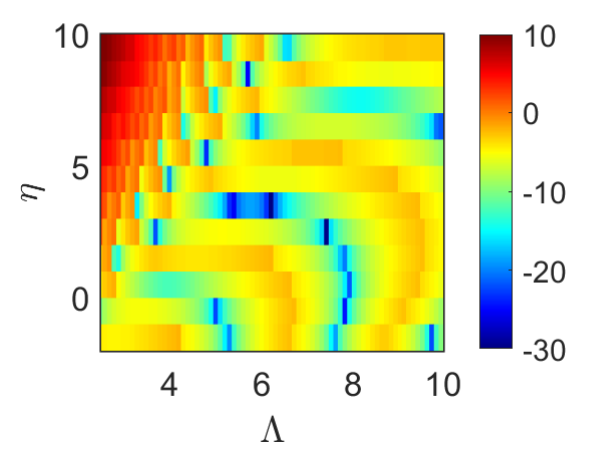}
    \caption{Difference between numerical and analytical half-life.}
\end{subfigure}
\hfill
\begin{subfigure}[b]{0.48\linewidth}
    \centering
    \includegraphics[width=\linewidth,trim=0.08cm 0.08cm 0.08cm 0.08cm,clip]{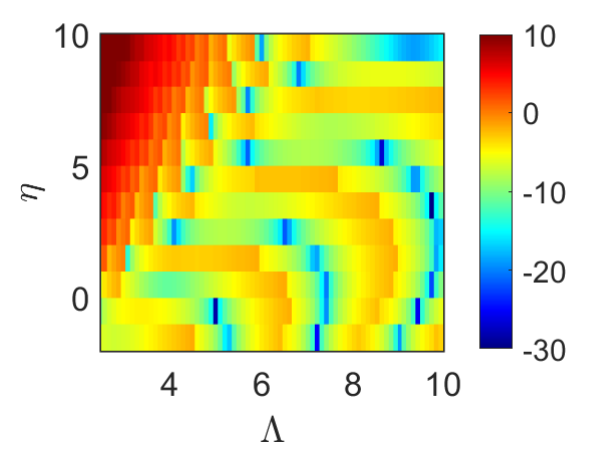}
    \caption{Difference between numerical and analytical lifetime.}
\end{subfigure}
\caption{Decay study for the precipitation covariance kernel $G_q$ over the $(\eta,\Lambda)$ plane. The first two panels compare the numerically evaluated lifetime with the analytical proxy derived from the reflectivity-weighted fall-spectrum width. The lower panels show the corresponding half-life and lifetime differences. The comparison confirms that the true precipitation kernel generally decays somewhat more slowly than the Gaussian proxy, especially at smaller $\Lambda$, but the relevant support remains short enough for the short-CPI simulated-data settings used in the paper.}
\label{fig:decay_study}
\end{figure*}

The numerical study, summarized in Fig.~\ref{fig:decay_study}, shows two consistent features. First, the precipitation kernel generally decays more slowly than this Gaussian proxy, especially for smaller $\Lambda$, which corresponds to larger characteristic drops. Second, for the range of $(\eta,\Lambda)$ values relevant to realistic rain DSDs, the substantially correlated part of the sequence still remains short enough to be covered by the coherent record lengths used in the simulated data. In practical terms, the lag support needed for the fall component is well below the values that would invalidate the short-CPI approximation in the simulated study considered here.

This decay study also explains the choice of quadrature order in the semianalytical evaluation of $G(q)$. The Gauss--Laguerre approximation must remain accurate over the lags that materially contribute to the PSD. In the present work $n=64$ is retained, which is sufficient over the decay interval that matters for the parameter ranges of interest, even though larger $n$ would be required to track the tail accurately at much longer lags. The practical conclusion is therefore straightforward: for the simulated-data configurations studied here, the decay of the precipitation kernel is fast enough that the finite-$N$ approximation and the retained quadrature order remain suitable for the retrieval problem addressed in the main text.

\section{CRB for the Latent Parameter Vector}
For completeness, the Cram\'er--Rao analysis is retained here for the latent vector
\begin{equation}
\bm{\theta}=[\mu_v,\sigma_v,\eta,\Lambda]^\top .
\end{equation}
Starting from \eqref{eq:ll}, the mean-spectrum derivative vector is
\begin{equation}
\mathbf{g}_i=\frac{\partial F(v_i;\bm{\theta})}{\partial \bm{\theta}},
\end{equation}
and the Fisher information matrix for $L$ incoherent spectra is approximated by
\begin{equation}
\mathbf{I}(\bm{\theta})
=
\sum_{i=1}^{N_f}
\frac{L}{\left(F(v_i;\bm{\theta})+\sigma_n^2\right)^2}
\mathbf{g}_i \mathbf{g}_i^\top.
\end{equation}
The latent-parameter CRB is then
\begin{equation}
\mathbf{C}_{\theta,\mathrm{CRB}}=\mathbf{I}^{-1}.
\end{equation}

The paper does not emphasize CRB curves for the latent parameters because those plots are not especially informative for the new scientific narrative. In finite samples the estimator can leave the local quadratic regime, and in that situation a latent-parameter CRB can look overly optimistic or simply irrelevant to the actual failure mode. The derivation is therefore retained for completeness, while the main paper focuses on the derived quantities and on finite-$L$ uncertainty trends that are more directly interpretable.

\section{Bounds and Finite-$L$ Variance Proxies for $D_m$ and $V_T$}
The bounds for $D_m$ and $V_T$ follow by first-order propagation from the covariance of $(\eta,\Lambda)$. Let
\begin{equation}
\mathbf{J}_{DV}
=
\begin{bmatrix}
\frac{\partial D_m}{\partial \mu_v} &
\frac{\partial D_m}{\partial \sigma_v} &
\frac{\partial D_m}{\partial \eta} &
\frac{\partial D_m}{\partial \Lambda} \\
\frac{\partial V_T}{\partial \mu_v} &
\frac{\partial V_T}{\partial \sigma_v} &
\frac{\partial V_T}{\partial \eta} &
\frac{\partial V_T}{\partial \Lambda}
\end{bmatrix},
\end{equation}
where the nonzero derivatives are
\begin{equation}
\frac{\partial D_m}{\partial \eta}=\frac{1}{\Lambda},
\qquad
\frac{\partial D_m}{\partial \Lambda}=-\frac{3.67+\eta}{\Lambda^2},
\end{equation}
and
\begin{equation}
\frac{\partial V_T}{\partial \eta}
=
\frac{C_2 \log\!\left(1+\frac{C_3}{\Lambda}\right)\left(1+\frac{C_3}{\Lambda}\right)^{-\eta-7}}{\sin\theta},
\end{equation}
\begin{equation}
\frac{\partial V_T}{\partial \Lambda}
=
-\frac{C_2(\eta+7)C_3}{\Lambda^2\sin\theta}
\left(1+\frac{C_3}{\Lambda}\right)^{-\eta-8}.
\end{equation}
The propagated covariance is then
\begin{equation}
\mathbf{C}_{DV}\approx \mathbf{J}_{DV}\,\mathbf{C}_{\theta}\,\mathbf{J}_{DV}^\top,
\end{equation}
where $\mathbf{C}_{\theta}$ may be taken either as the CRB matrix or, as done in the \mbox{$L$-sweep}, as a finite-$L$ Fisher covariance proxy evaluated at the truth model.

This is the quantity used for the theoretical standard-deviation overlays in Fig.~\ref{fig:L_sweep}. The purpose of those overlays is not to claim exact lower bounds for the finite-sample SPLIT estimator, but rather to show that the empirical trend with $L$ is consistent with the information increase predicted by the forward model. This appendix therefore plays a supporting role: it connects the simulation trends to a transparent uncertainty calculation without forcing the paper back into latent-parameter performance claims.

\bibliographystyle{unsrt}
\bibliography{references}

\end{document}